# High pressure study on LaFeAs($O_{1-x}F_x$) and LaFeAsO$_\delta$ with different Tc

Wei Yi, Chao Zhang, Liling Sun[*], Zhi-an Ren, Wei Lu, Xiaoli Dong,
Zhengcai Li, Guangcan Che, Jie Yang, Xiaoli Shen,
Xi Dai, Zhong Fang, Fang Zhou, Zhongxian Zhao[*]

Institute of Physics and Beijing National Laboratory for Condensed Matter Physics, Chinese Academy of Sciences, Beijing 100190, P. R. China

Abstract

We report studies on pressure dependence of superconducting transition temperature (Tc) of LaFeAsO$_\delta$ ($\delta<0.3$), LaFeAs($O_{0.5}F_{0.5}$) and LaFeAs($O_{0.89}F_{0.11}$) samples. *In-situ* resistance measurements under high pressure showed that the Tc of these three compounds increases with pressure initially, reaches a maximum value and then decreases with further increasing pressure, although the Tc at ambient pressure are different. The onset Tc of LaFeAsO$_\delta$ is ~50 K at 1.5 GPa, which is the highest record in La-based oxypnictide system. The significant change in Tc induced by pressure is attributed to the orbital degeneracy and the electron density of state at the Fermi level.

Corresponding author:
llsun@aphy.iphy.ac.cn
zhxzhao@aphy.iphy.ac.cn

Since the Fe-containing superconductor with transition temperature of 26 K was discovered [1], a large body of experimental and theoretical work on this subject has been carried out. It has been found that the critical transition temperature (Tc) was increased significantly by replacing La with Ce, Pr, Nd, Sm, Gd and established the record of 55 K at ambient pressure to date [2-7]. Resistance and magnetization measurements under high pressure for optimal oxygen doped $SmFeAsO_{0.85}$ and $NdFeAsO_{0.85}$ sample indicated that Tc decreases with pressure [8, 9]. The reason for the negative pressure effect on Tc is due to that pressure drives a decrease of the electron density of states at the Fermi level. For $LaFeAs(O_{0.89}F_{0.11})$ compound, high pressure studies carried out by resistance and magnetic measurements exhibited that the Tc increases with pressure in the low pressure region [10, 11], passes the peak and then decreases with further increasing pressure [11].

It has been believed that the Fe-As layer is responsible for the superconductivity, while the Re-O (Re= Ce, Pr, Nd, Sm etc) layer provides charge carriers. The shrinkage of the lattice induced by internal or external pressure could increase Tc. In this paper, we report high-pressure studies on superconductivities of $LaFeAs(O_{1-x}F_x)$ and $LaFeAsO_\delta$ sample via *in-situ* resistance measurements in a diamond anvil cell and establish pressure (P) and transition temperature (Tc) diagram for each sample. Fitting to the data measured under high pressure for different samples, and extrapolating the Tc to ambient pressure found that the Tc of $LaFeAs(O_{1-x}F_x)$ and $LaFeAsO_\delta$ could be achieved around 50-55 K under pressure.

The sample 1, $LaFeAs(O_{0.89}F_{0.11})$, was prepared at ambient pressure (AP) as described in Ref. [1, 10]. Its onset Tc is about 26 K. The sample 2, LaFeAs($O_{0.5}F_{0.5}$), was synthesized under high pressure and high temperature (HPHT). DC magnetization and resistance experiments at ambient pressure displayed that their onset Tc's are 35 K and 37.5 K respectively [12]. The sample 3, LaFeAs$O_\delta$, without F-doping but with oxygen vacancy, was also synthesized under HPHT by solid state reaction method which has been reported in Ref. [3]. The nominal delta of the sample 3 is less than 0.3. The magnetization measurements under 1 Oe after zero-field cooling (ZFC) and field cooling (FC) for the sample 3 were carried out using the Quantum Design Magnetic Property Measurement System (MPMS-XL1). The obvious diamagnetic behavior shown in Fig. 1(a) demonstrates that there exists genuine bulk superconductivity. The onset critical transition temperature determined by ZFC measurement is 38 K, as shown in Fig. 1(b). The lattice parameter (*a*) dependence of Tc for those samples agrees with the phase diagram of Tc vs *a* reported in our previous work [13]. Characterization by powder X-ray diffraction with Cu-Kα radiation verifies that the main phase of each sample is 1111 phase, and impurity phases are FeAs, $Fe_2O_3$, and $FeAs_2$, as shown in Fig. 1(c). These impurity phases are not superconducting at the temperature and pressure range considered in this study.

High-pressure resistance measurements of these samples were carried out in a diamond anvil cell made of Be-Cu alloy. The four-standard-probe technique was adopted in the experiments, in which 2μm-thick platinum (Pt) plates were used as electrodes and insulated from a rhenium gasket by a thin layer of the mixture of cubic boron nitride and epoxy. Pressure determined by ruby fluorescence method at room temperature [14] was calibrated through the measurements of Tc change of lead as a function of pressure [15]. The superconductivity transition of the three samples at each loading point was detected using the CSW-71 refrigerator system.

Fig. 2 shows the representative $R/R_{80K}$-T (resistivity normalized at 80K) results for LaFeAs($O_{0.5}F_{0.5}$) and LaFeAs($O_{0.89}F_{0.11}$) sample at different pressures. It can be seen that the onset Tc of LaFeAs($O_{0.5}F_{0.5}$) sample prepared at HPHT and LaFeAs($O_{0.89}F_{0.11}$) sample prepared at ambient is 37.5K and 26.3K at ambient pressure, respectively. With increasing pressure, a dome-shaped diagram of pressure dependence of onset Tc was observed in the sample 1 and the sample 2. The pressure effect on Tc for the sample 2 is in good agreement with the data reported in Ref. [11], whereas the P-Tc behavior of the sample 1 is roughly consistent with the results of Ref. [16].

Fig. 3(a) shows the representative results of electrical resistance ratio $R/R_{80K}$ of the sample 3 (LaFeAsO$_\delta$) as a function of temperature at different pressures. At ambient pressure, the temperature dependence of resistivity clearly shows superconducting transition at 43 K where the resistance of the sample begins to decrease. This is the highest onset Tc among the La-based superconductor achieved at ambient pressure so far. In this high pressure study, the zero resistance was failed to observe because of the micro-cracks introduced during non-hydrostatic pressuring that has been explained in our previous report [8]. The pressure dependence of the superconducting transition temperature was plotted in Fig. 3(b). Here the onset superconducting transition temperature (Tco) was determined as the starting temperature where dR/dT raises (i.e., where resistance starts to decrease because superconductivity takes part in), as illustrated in inset of Fig. 3(a). Initially, the onset Tc increases rapidly with pressure up to ~1.5 GPa, in which the pressure coefficient dTc/dP is 7.5 K/GPa. The Tc passes the maximum and then decreases monotonously at pressure above 1.5 GPa. The pressure coefficient dTc/dP for the pressure from 1.5 GPa to 13 GPa is about -1.2 K/GPa, showing a dome-like P-Tco diagram. Although the onset Tc of the LaFeAsO$_\delta$ sample is higher than that of the LaFeAs(O$_{0.89}$F$_{0.11}$) sample [11], the behavior of pressure-induced changes in Tc appears quite same.

The Tc change as a function of pressure for three samples was plotted in Fig. 4. Fitting to the data in the 'high' pressure regime for the three samples, and extrapolating the fitting to the ambient pressure, it was found that the Tc at ambient pressure could be achieved in the temperature range of 50-55 K.

The effects of pressure on the electronic structure in LaFeAs(O$_{1-x}$F$_x$) have been studied by *ab initial* calculation in Ref. [17]. For such a multi-orbital system, we need two characteristic quantities to describe the electronic structure near the Fermi surfaces (FS), namely, the orbital degeneracy and density of states (DOS) at the FS. The results in Ref. [17] and Ref. [8] indicate that in the linear region the former increases while the latter decreases with the pressure. Since both the high DOS and orbital degeneracy will enlarge the effective phase space for pairing, we believe the non-monotonic behavior of the pressure effect here is due to the integrative effect of the increment of orbital degeneracy and the decrement of DOS at the FS.

Another reason for the P-Tc behavior in La-based oxypnictide system may be associated with pressure-induced phase transition. To clarify this inference, *in-situ* high-pressure x-ray diffraction measurements are needed.

In summary, the superconductivities of La-based oxypnictide system with F-doping and oxygen vacancy doping were studied under high pressure via resistance measurements in a diamond anvil cell. A common feature of P-Tc behaviors in LaFeAs($O_{1-x}F_x$) and LaFeAsO$_\delta$ ($\delta<0.3$) has been observed as that Tc increases with pressure initially, passes the peak and then decreases with further increasing pressure. The Tc in LaFeAsO$_\delta$ is ~50 K at 1.5 GPa, which is the highest record in La-based oxypnictide system. The Tc response to the pressure for three samples is related to the orbital degeneracy and the electron density of state at the Fermi level.

Acknowledgements

We wish to thank the National Science Foundation of China for its support of this research through Grant No. 50571111, 10874230 and 10734120. This work was also supported by the Ministry of Science and Technology of China (2005CB724400, 2006CB601001 and 2007CB925002). We also acknowledge the support from EU under the project CoMePhS.

Figure captions:

Fig. 1. (a) Temperature dependence of the magnetization measured under 1Oe for the LaFeAsO$_\delta$ sample. (b) Differential curve of the ZFC magnetization (dM/dT) shown in (a). (c) X-ray diffraction patterns of LaFeAs(O$_{0.5}$F$_{0.5}$), LaFeAs(O$_{0.89}$F$_{0.11}$) and LaFeAsO$_\delta$ samples.

Fig. 2. (a) Representative electrical resistance of the LaFeAs(O$_{0.5}$F$_{0.5}$) sample vs. temperature at different pressures. (b) Electrical-resistance of the LaFeAs(O$_{0.89}$F$_{0.11}$) sample vs. temperature at different pressures.

Fig. 3. (a) Electrical resistance of the LaFeAsO$_\delta$ sameple as a function of temperature at different pressures. The inset shows dR/dT curve of the LaFeAsO$_\delta$ sample at ambient pressure, by which the onset Tco is determined. (b) Pressure dependence of Tc of the LaFeAsO$_\delta$ sameple for different runs.

Fig. 4: Diagram of superconducting Tc vs. pressure for LaFeAsO$_\delta$, LaFeAs(O$_{0.5}$F$_{0.5}$) and LaFeAs(O$_{0.89}$F$_{0.11}$) sample.

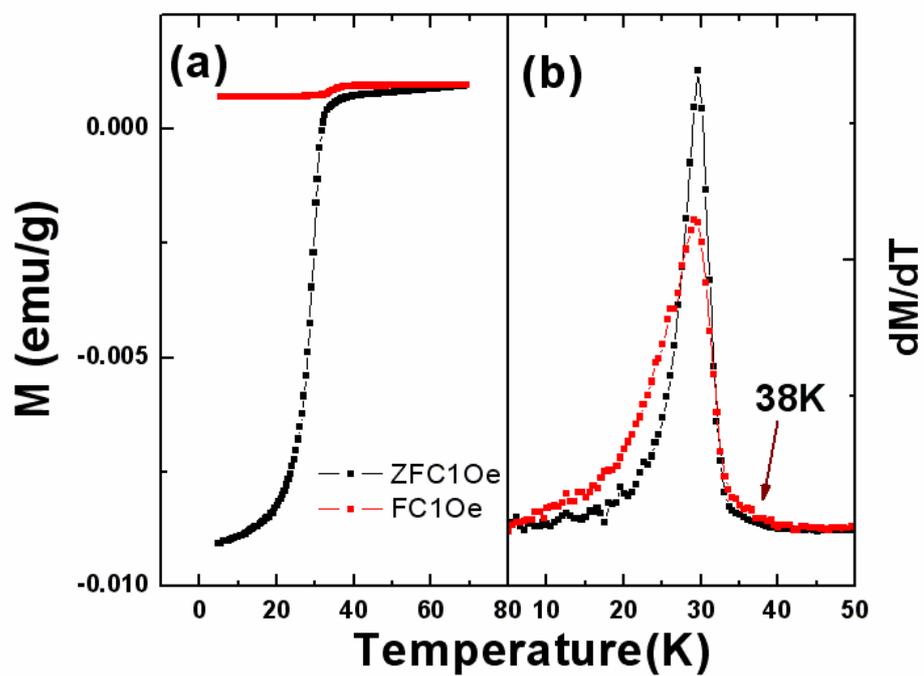

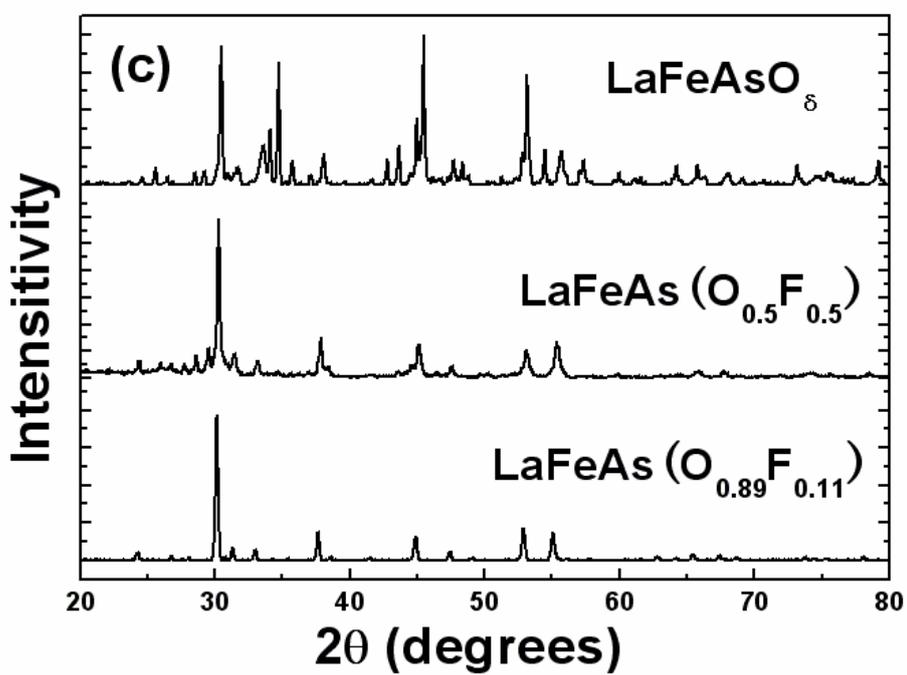

Fig. 1

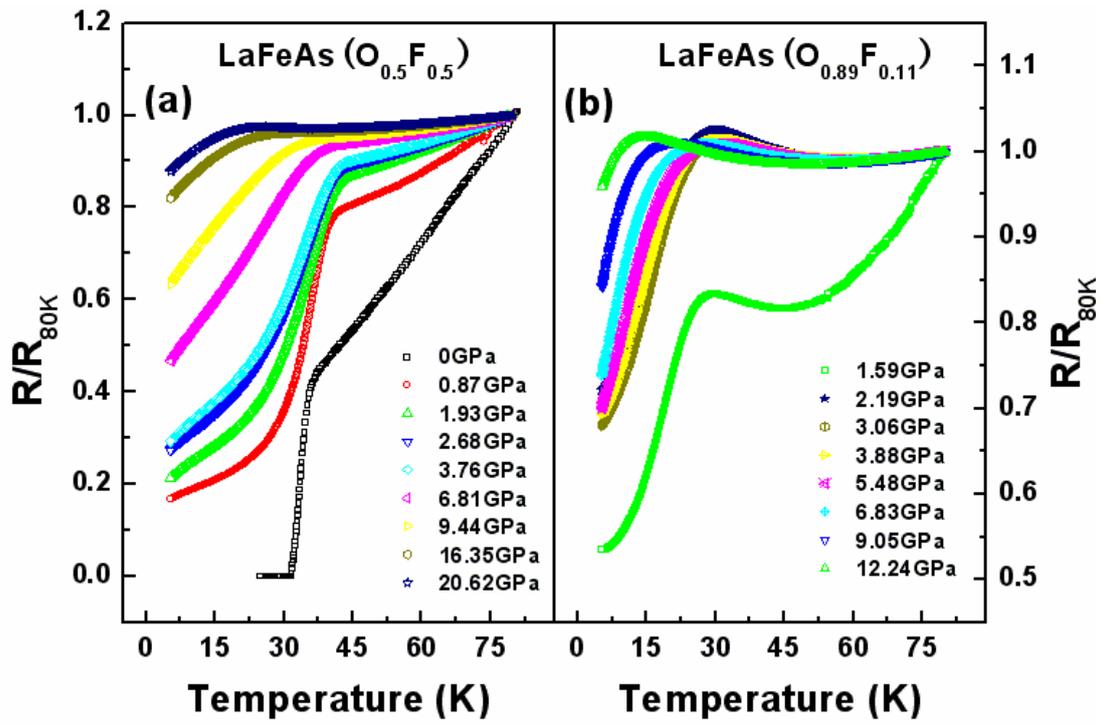

Fig. 2

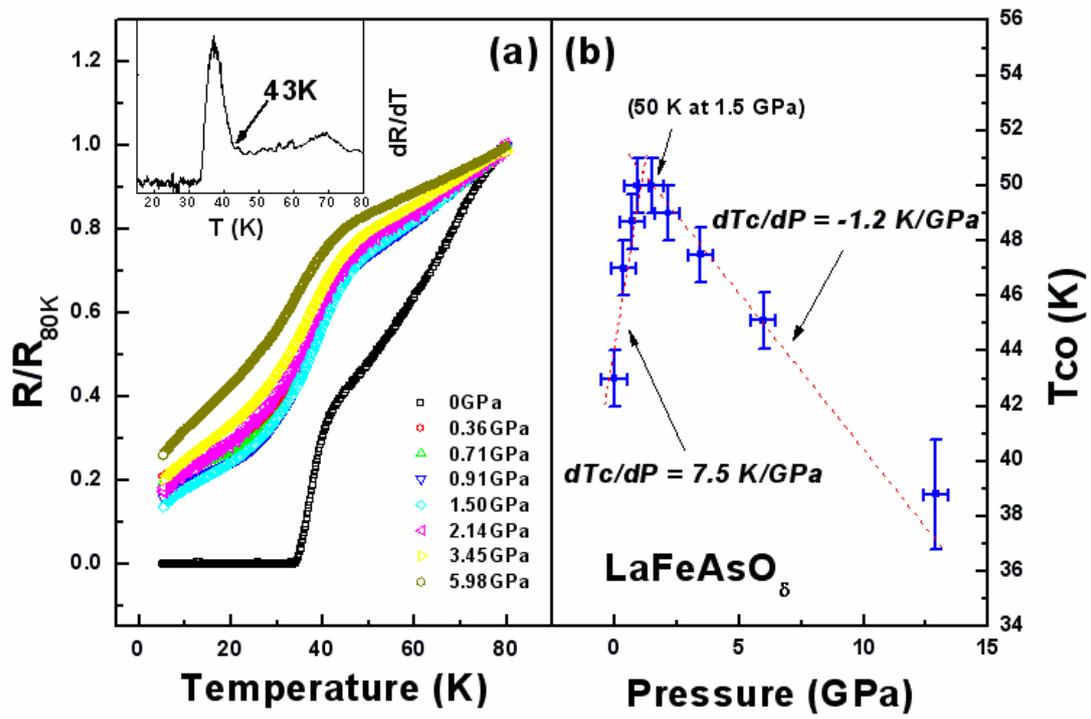

Fig. 3

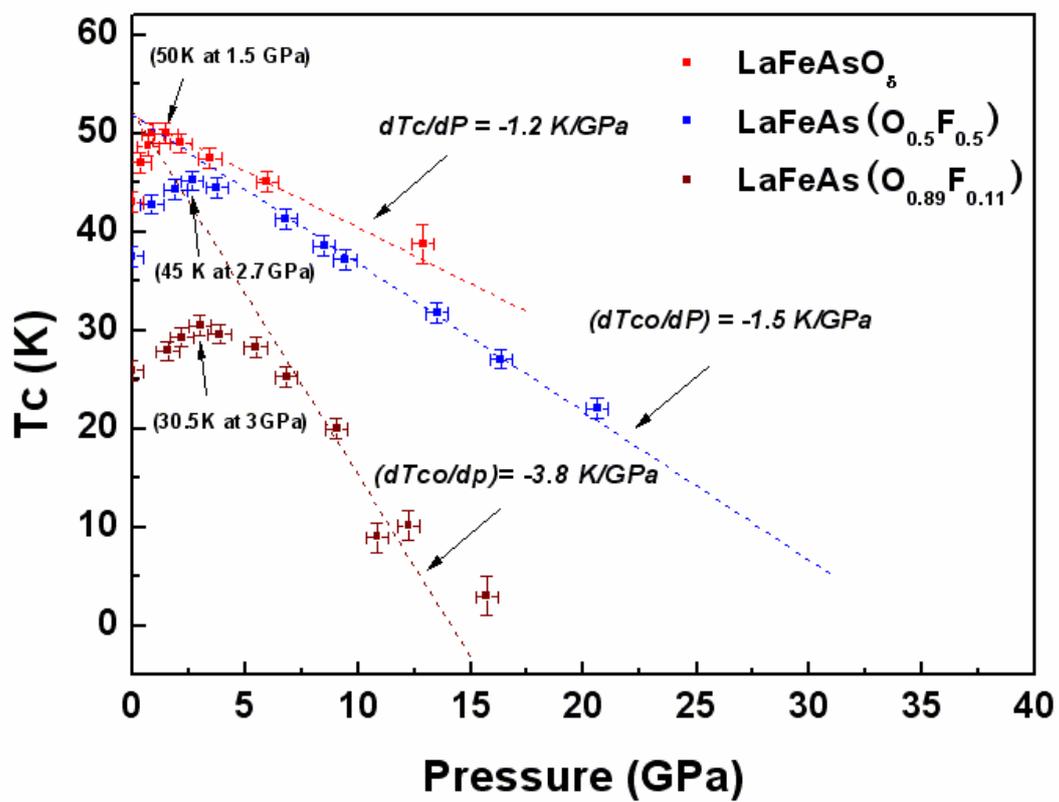

Fig.4